\documentclass[12pt]{article}
\input xy
\xyoption{all}
\input epsf.tex
\usepackage{pstricks}

\usepackage{graphicx}
\usepackage{mathrsfs}
\usepackage{cancel}
\usepackage{amssymb,amsmath}
\def\harr#1#2{\smash{\mathop{\hbox to .3in{\rightarrowfill}}
 \limits^{\scriptstyle#1}_{\scriptstyle#2}}}

\def\s2{\frac{1}{\sqrt2}}

\def\be{\begin{equation}}
\def\ee{\end{equation}}
\def\beqa{\begin{eqnarray}}
\def\eeqa{\end{eqnarray}}

\def\Dsl{\,\raise.15ex\hbox{/}\mkern-13.5mu D} %can be subscripted
\def\d3{d^3}

\def\IZ{\mathbb{Z}}

\begin{document}

\begin{center}
\Large{\bf Gauged WZW Models Via Equivariant Cohomology}\\
\vspace{1cm}

\large Hugo Garc\'{\i}a-Compe\'an\footnote{e-mail address: {\tt
compean@fis.cinvestav.mx}}, Pablo Paniagua\footnote{e-mail address: {\tt
ppaniagua@fis.cinvestav.mx}}
\\
[2mm]
{\small \em Departamento de F\'{\i}sica, Centro de
Investigaci\'on y de
Estudios Avanzados del IPN}\\
{\small\em P.O. Box 14-740, 07000 M\'exico D.F., M\'exico}\\
[4mm]

\vspace*{2cm}
\small{\bf Abstract}
\end{center}
The problem of computing systematically the gauge invariant extension of WZW term
through equivariant cohomology is addressed. The analysis done by Witten in the
two-dimensional case is extended to the four-dimensional ones. While Cartan's model
is used to find the anomaly cancelation condition. It is shown that
the Weil model is more appropriated to find the gauge invariant extension of the WZW term.
In the process we point out that Weil's and Cartan's models are also useful to stress some connections with the abelian anomaly.
\begin{center}
\begin{minipage}[h]{14.0cm} {}
\end{minipage}
\end{center}

\bigskip
\bigskip

\date{\today}
%\leftline{CINVESTAV-FIS/02-027}
%\leftline{\tt hep-th/yymmnnn}

\vspace{3cm}

\leftline{August, 2010}
\newpage

%%%%%%%%%%%%%%%%%%%%%%%%%%%%%%%%%%%%%%%%%%%%%%%%%%%%%%%%%%%%%%%%%%%%%%%%%%%%%%%%%%%%%%
%%%%%%%%%%%%%%%%%%%%%%%%%%%%%%%%%%%%%%%%%%%%%%%%%%%%%%%%%%%%%%%%%%%%%%%%%%%%%%%%%%%%%%
%%%%%%%%%%%%%%%%%%%%%%%%%%%%%%%%%%%%%%%%%%%%%%%%%%%%%%%%%%%%%%%%%%%%%%%%%%%%%%%%%%%%%%

Wess-Zumino-Witten term $\Gamma_{WZW}$ \cite{WessZumino71,Witten:1983tw} contains a great
deal of important information for some models in physics. It is well
known that them describes some decays, for
instance: the celebrated $\pi^0 \to 2 \gamma$ and $K\overline{K} \to 3 \pi$,
which are driven by the presence (or absence) of chiral anomalies
\cite{Treiman:1986ep}. In this same context the WZW term was used to study
the baryons as solitons \cite{Witten:1983tx}.

In the present paper we will be interested in the low energy effective action of
a theory at higher energy describing Yang-Mills fields coupled to $N_f$ species of quarks (flavors).
These effective actions are precisely the gauged $\Gamma_{WZW}$ actions which are invariant under an
anomaly free
subgroup $H$ of the original theory $U(N_f)_L \times U(N_f)_R$.
Recently the WZW terms have been studied in the context of the topological interactions that follows a composite or little Higgs \cite{Hill:2007nz}. Moreover, more recently in Ref. \cite{Harvey:2007ca} it was found that still is
possible to gauge specific non-abelian groups, as the Standard Model group $SU(2)_L \times
U(1)_Y$ with the condition of adding local counterterms where some novel interactions were studied. Furthermore in Ref. \cite{Hill:2009wp} a topological derivation of the WZW term for the Standard Model Higgs field is obtained through an interesting symmetry breaking reduction.

More recently  in Ref. \cite{Witten:1991mm} Witten realizes that in a gauged WZW model
in two dimensions,
the free anomaly condition can be expressed as the condition of the
existence of a closed equivariant extension of the Cartan 3-form associated to the WZW term.
Moreover he was able to find an invariant gauge version of the WZW action in this context. Later some generalizations where implemented through the consideration  of some vanishing theorems in equivariant cohomology which ensure the absence of obstructions to gauge the WZ term, implying some restrictions on the target space and symmetry groups \cite{FigueroaO'Farrill:1994dj}.

Equivariant cohomology is a mathematical structure very useful in mathematics to
understand problems in symplectic geometry, see for instance
\cite{Atiyah:1984px,Mathai:1986tc,BottTuequiv,Wu:1993iia}. In particular in Ref. \cite{BottTuequiv} Bott and Tu
gave a more al suitable description of the equivariant cohomology in order to provide a closest relation
between their geometric and topological approaches. Precisely their view is that will be used in the present
paper to reexamine the description of the chiral anomalies and the systematic construction of the
gauge invariant extension of the WZW action. Conventions and notation are also taken from \cite{BottTuequiv}.
In the present paper we find some suitable fourth-dimensional implementation of this construction\footnote{The construction can be also carried over in any higher even dimension, however for the sake of simplicity we prefer to restrict ourselves to four dimensions.}.
The procedure corresponds with finding
some five-dimensional cocycles in the equivariant cohomology of
the underlying target space. In the process we will able to find a systematic way of obtaining the gauge
invariant extension for the WZW actions. Thus we confirm explicitly Witten's analysis
(done specifically for the two-dimensional case) now for the fourth-dimensional
case. Let start with the WZW model in four dimension given by
\begin{equation}\label{wzwa}
    {I}=  \frac{1}{4} F_{\pi}^{2} \int d^{4}x {\rm Tr}\big(\partial_{\mu}U \partial^{\mu}U^{-1}\big)
 + N \Gamma_{WZW} + {\rm higher \ order \ terms}
\end{equation}
where $F_{\pi}\approx 93 MeV$ for QCD is the pion decay constant and
\begin{equation}\label{}
    \Gamma_{WZW} = \frac{-iN_{c}}{2\pi^{2} \times 5!} \int d\Sigma^{ijklm} (U^{-1}\partial_{i}U)(U^{-1}\partial_{j}U)(U^{-1}\partial_{k}U)(U^{-1}
    \partial_{l}U)(U^{-1}\partial_{m}U).
\end{equation}
Here $d\Sigma^{ijklm}$ is the volume element of a five-dimensional disc. The WZW term is of topological nature and contributes with the appropriate phenomenological symmetries of the theory\footnote{ For instance, for QCD these symmetries are: $P=P_{0}(-1)^{N_B}$, where $P_{0}$ is the naive parity operation ($x\leftrightarrow -x$, $t \leftrightarrow t$, $U \leftrightarrow U$) and $(-1)^{N_B}$ is the operation $U \leftrightarrow U^{-1}$ that counts modulo two the number of bosons $N_{B}$.}.

It is convenient to rewrite the WZW action (\ref{wzwa}) as
\begin{equation}
\label{3}
I= - \frac{\Lambda^{2}}{4} \int_{M}  {\rm Tr} \big( g^{\ast} (\theta) \wedge * (g^{\ast}(\theta))\big)
-ic_4 \int_{\bf D} g^{\ast} \omega _{5},
\end{equation}
where $c_4$ is a constant, $\theta := g^{-1}dg \in \Omega^{1}(G,\mathfrak{g})$ is the Maurer-Cartan form, $\omega_5 = {\rm Tr} (\theta^5) \in H^5(G,\IZ)$ is the Cartan 5-form, $g: M \to G$, with $M$ being the spacetime manifold (of dimension $D$) that we will take closed and
without boundary. $G$ is the target space which is a compact connected Lie group of rank $n \geq 5$. We consider $M$ to be the one-point compactification of Euclidean space such that $M$ is the boundary of a 5-disk ${\bf D}$ whenever $\pi_4(G)=0$. Action (\ref{3}) is invariant under the group $G_L \times G_R$ whose transformation is given by: $\Psi: (G_L \times G_R) \times G \to G$, $((a,b),g) \mapsto agb^{-1}$, where $G_L$ and $G_R$ are two copies of $G$.

In general the WZW term $\Gamma_{WZW}$ does no have a gauge invariant extension under local gauge
transformations associated to $G_L \times G_R$. Then one ask what conditions must be satisfied in order a gauge invariant extension of the WZW term does exist. Its answer is hard to respond for the general case, but it depends of the existence of an anomaly-free subgroup $H \subset G_{L} \times G_{R}$ whose generators should satisfy an algebraic condition. Let $\{T_{1,L},T_{2,L},...,T_{dim \; G,L}\}$ and $\{T_{1,R},T_{2,R},...,T_{dim \; G,R}\}$ be basis sets for $\mathfrak{g}_{L}$  and $\mathfrak{g}_{R}$ respectively. Then if we have a subgroup $H$ of $G_{L} \times G_{R}$, its Lie algebra $\mathfrak{h}$ of $H$ have generators $K_{a}$ which are linear combinations of $T_{a,L}$ and $T_{a,R}$ for some subset of indices $a$ and zero for the rest. In general the subgroup $H$ is defined by the condition of absence of anomalies.

In $D=2$ dimensions, the non-abelian chiral gauge extension of the WZW term exist if
\begin{equation}
\label{10}
D_{ab}= \frac{1}{2\pi} \big[ {\rm Tr}(T_{a,L}T_{b,L}) - {\rm Tr}(T_{a,R}T_{b,R})\big]=0,
\end{equation}
which is the condition for absence of anomalies and the subgroup $H$ have generators satisfying this condition.

Now in $D=4$ dimensions, if we have a theory with local symmetry $G_{L} \times G_{R}$, then the theory is completely free of anomalies if and only if all the triangle graph anomalies are absent. The triangle diagram of chiral fermions has an anomaly that is proportional to
\begin{equation}\label{104D}
D_{abc}=   {\rm Tr}(\{T_{a,L},T_{b,L}\}T_{c,L}) - {\rm Tr}(\{T_{a,R},T_{b,R}\}T_{c,R}).
\end{equation}
Then the chiral anomaly vanishes if the term $D_{abd}$ is zero. Moreover for the case of gauge symmetries this requirement this is the only way of saving renormalizability and unitarity of the fundamental theory.

Going back to the two-dimensional case we assume that $M \simeq S^{2} $ which is boundary of the
disk ${\bf D}^{3}$. Thus we have a WZW term determined by the Cartan's 3-form
\begin{equation}
\label{}
\omega_{3} = \frac{1}{12 \pi} {\rm Tr}  [  g^{-1}dg  \wedge g^{-1}dg  \wedge  g^{-1}dg],
\end{equation}
where $ \omega_{3} \in H^{3}(G)$ and $G$ is compact connected and $\pi_{2}(G)=0$.

We are interested in finding a closed equivariant extension of $\omega$. It is easy to see that this
extension is of the form
\begin{equation}
\label{}
\widetilde{ \omega}_3 = \omega _3 + \sum_{a} \phi^{a} \lambda_{a},
\end{equation}
where $\phi_a$'s are variables of degree 2 generating the symmetric algebra of the Weil algebra ${\cal W}(\mathfrak{g})$,
$\widetilde{ \omega}_3 \in \Omega^{3}_{H}(G)$, $H$ is a subgroup
of $G_L \times G_R$ to be determined. The equations to be satisfied for such a closed equivariant extension are
\begin{equation}
\label{2decf1}
i_{{X_{a}}}  \omega_{3} = d \lambda _{a},
\end{equation}
\begin{equation}
\label{2decf2}
\sum_{a,b} \phi^{a} \phi^{b} i_{{X_{a}}} \lambda _{b} = 0.
\end{equation}
It is remarkable that, under the Chern-Weil homomorphism and the pull-back to spacetime,
the lhs of eq. (\ref{2decf2}) can be identified with the abelian anomaly in four dimensions.

Recalling that the symmetry $G_L\times G_R$ of the theory is encoded in the nonlinear action $\Psi$ of
the chiral group on $G$. On the other hand, we note that, under this action, we have the fundamental
vectors: $[X_{a}]_{g} = T_{a,L} g - g T_{a,R} \equiv {(T_{a,L},T_{a,R})}_{g}$.

We solve eq. (\ref{2decf1}) in the context of equivariant cohomology and we get
$\lambda_{a} = - \frac{1}{4\pi} {\rm Tr}[ T_{a,L} dg g^{-1} + T_{a,R} g^{-1} dg ]$. It remains to
solve equation eq. (\ref{2decf2}), whose
solution is $i_{{X_{a}}} \lambda _{b} + i_{{X_{b}}} \lambda _{a} =0$ due the
algebra generated by the ${\phi}$'s is a symmetric algebra.
Then we finally obtain $i_{{X_{a}}} \lambda _{b} + i_{{X_{b}}} \lambda _{a} =\frac{1}{2\pi} {\rm Tr}[ T_{a,L}T_{b,L} -T_{a,R}T_{b,R} ]=0.$ Thus if we require the existence of the closed equivariant extension of
$\omega_{3}$, it must satisfy the condition: $D_{ab}=0$ \cite{Witten:1991mm}. This is precisely the chiral anomaly cancelation condition for a gauge theory in 2 dimensions. Moreover this condition also define the anomaly free subgroup $H \subset G_L \times G_R$. Summarizing we have accomplished to lift the Cartan 3-form $\omega_{3} \in H^{3}(G)$ i.e. we find $\widetilde{\omega}_{3} \in H_{H}^{3}(G)$.

For the theory in four dimensions the WZW term does not has a gauge invariant extension unless we restrict ourselves to work with an anomaly-free subgroup $F$ of $G_{L} \times G_{R}$. Of course the WZW effective lagrangian is invariant under the usual action of $G_{L} \times G_{R}$ on $G$. Then the condition for absence of anomalies from eq. (\ref{104D}) is $D_{abc}=0$. This statement is equivalent to the statement that the class in $H^{5}(G,\mathbb{Z})$, represented by $\omega_{5}$, has an extension in $H_{F}^{5}(G)$. To be more precise we note that an equivariant extension is given by
\begin{equation}
\label{}
\widetilde{\omega}_{5} = \omega_{5} + \sum_{a}\phi^{a} \beta_{a} + \sum_{a,b} \phi^{a} \phi^{b} \alpha_{ab},
\end{equation}
where $ \widetilde{\omega}_{5} \in \Omega^{5}_{F}(G)$ and $F$ is the anomaly-free subgroup of $G_L\times G_R$. Consequently the equations that must fulfill $\omega_{5}$, $\beta_{a}$ and $\alpha_{ab}$, are contained in the following expression

\begin{equation}
\label{acc}
J= \sum_{a} \phi^{a} (-d \beta_{a} + i_{{X_{a}}} \omega_{5}) + \sum_{a,b} \phi^{a} \phi^{b} (d \alpha_{ab} - i_{{X_{a}}} \beta_{b}) + \sum_{a,b,c} \phi^{a} \phi^{b} \phi^{c} i_{{X_{c}}} \alpha_{ab}=0.
\end{equation}
In order to solve these equations we must take into account that the products of variables $\phi^{a}$ are symmetric and consequently their solution is given by: $\beta_{a}=- 5 $ ${\rm Tr} T_{a,L} (dg g^{-1})^{3}$ $+ T_{a,R} (g^{-1}dg)^{3}]$ and $\alpha_{ab}=  5 \;\mathrm{STr}\big[ 2T_{a,L}T_{b,L} \; dg g^{-1} + 2T_{a,R}T_{b,R}\; g^{-1}dg + T_{a,R} g^{-1} T_{b,L} dg - T_{a,L} g T_{b,R} dg^{-1} \big]$, where STr is the symmetrized trace. Details of the computation implies that the first and the second terms of the rhs of (\ref{acc}) vanish. Then it remains to find the last term and this yields
\begin{equation}
J= \sum_{a,b,c} \phi^{a} \phi^{b} \phi^{c} 5 {\rm Tr} [\{ T_{a,L},T_{b,L} \} T_{c,L} -
\{T_{a,R},T_{b,R}\} T_{c,R}]=0.
\label{jota}
\end{equation}
Notice  that under the Chern-Weil homomorphism and the pull-back to spacetime, the lhs of eq. (\ref{jota}) is precisely the abelian anomaly in six dimensions. The last equality can be obtained rearranging in an appropriate way the indices ${a,b,c}$ and taking into account that the products of variables $\phi^{a}\phi^{b}\phi^{c}$ are symmetric. Then taking the condition that $\widetilde{\omega}_5$ is a closed equivariant extension  we get the anomaly cancelation condition in a non-abelian chiral theory
\begin{equation}
\label{}
{\rm Tr}[ \{ T_{a,L},T_{b,L} \} T_{c,L}] = {\rm Tr}[ \{T_{a,R},T_{b,R}\} T_{c,R} ].
\end{equation}
This condition closed fines precisely the anomaly free subgroup $F \subset G_L\times G_R$ and give us the existence of the equivariant extension of the Cartan's form $\omega_5$.

Equivariant cohomology has been useful to understand the cancelation anomaly condition through the  closed equivariant extension form in the Cartan model. Now we will show that it also describes the gauge invariant
extension of the WZW term. In particular, the Weil model will be relevant to make this procedure by using the Chern-Weil homomorphism. Starting from $\widetilde{\omega}_5$,  under the
Mathai-Quillen isomorphism we get

\begin{eqnarray}
\label{weil ext 5 theta}
\nonumber
\widetilde{\omega}_{5}^{W} &=& a + \sum_{i} \Theta^{i} a_{i}
+ \frac{1}{2} \sum_{i,j} \Theta^{i}\Theta^{j} a_{ij}
+\frac{1}{6}\sum_{i,j,k}\Theta^{i}\Theta^{j}\Theta^{k}a_{ijk} \\
&+& \frac{1}{24} \sum_{i,j,k,l}\Theta^{i}\Theta^{j}\Theta^{k}\Theta^{l} a_{ijkl}
+ \frac{1}{120} \sum_{i,j,k,l,m}\Theta^{i}\Theta^{j}\Theta^{k}\Theta^{l}\Theta^{m} a_{ijklm},
\end{eqnarray}
where $\Theta^i$ are the generators of the anti-symmetric part of the Weil algebra, the coefficients are explicitly given in terms of all the non-vanishing contractions of
$\omega_{5}$, $\beta_{a}$ y $\alpha_{ab}$ with the fundamental vector fields associated to $\Psi$. For instance we have
$$
a_{i}= - i_{{X_{i}}}a, \;\;\; \ \ \ a_{ij}= -i_{{X_{i}}} i_{{X_{j}}}a, \;\;\; \ \  \
a_{ijk}= i_{{X_{i}}} i_{{X_{j}}} i_{{X_{k}}} a,
$$
\begin{equation}
 a_{ijkl}= i_{{X_{i}}} i_{{X_{j}}} i_{{X_{k}}} i_{{X_{l}}}a, \ \ \ \ \ \ \ \  a_{ijklm}= - i_{{X_{i}}} i_{{X_{j}}} i_{{X_{k}}} i_{{X_{l}}} i_{{X_{m}}}a,
\end{equation}
where  $a= \widetilde{\omega}_{5}$. Thus, after a long calculation we find that in the Weil formalism
the closed equivariant extension of $\omega_{5}$ reads
\begin{equation}
 \omega_{5}^{W}= \mathrm{Tr} (g^{-1} dg)^{5} + d \Xi,
\end{equation}
where
\begin{eqnarray}
\nonumber
   \Xi&=& -5\;\Theta^{i} \mathrm{Tr}[T_{i,L}\; (dg g^{-1})^{3} + T_{i,R}\; (g^{-1} dg)^{3} ] \\ \nonumber
   &+& 5\; d\Theta^{i}\cdot \Theta^{j} \mathrm{Tr}[(T_{i,L} T_{j,L}+ T_{j,L}T_{i,L})dg g^{-1}]\\ \nonumber
   &+& 5\; d\Theta^{i}\cdot \Theta^{j}\mathrm{Tr}[(T_{i,R}T_{j,R}+T_{j,R}T_{i,R})g^{-1}dg]\\ \nonumber
   &+& 5\; d\Theta^{i}\cdot \Theta^{j}\mathrm{Tr}[T_{i,L}dg T_{j,R}g^{-1}-T_{i,R} dg^{-1} T_{j,L}g]\\ \nonumber
   &-& 5\; \Theta^{i}\Theta^{j} \mathrm{Tr}[T_{i,R}g^{-1}T_{j,L}g (g^{-1} dg)^{2}- T_{i,L}g T_{j,R}g^{-1}(dg g^{-1})^{2}]\\ \nonumber
   &+&\frac{5}{2}\; \Theta^{i}\Theta^{j}\mathrm{Tr}[T_{i,L}dgg^{-1}\wedge T_{j,L}dg g^{-1} -T_{i,R}g^{-1}dg \wedge T_{j,R}g^{-1}dg] \\ \nonumber
   &+& 5\;  \Theta^{i}\Theta^{j}\Theta^{k} \mathrm{Tr}[T_{i,L}T_{j,L}T_{k,L}dg g^{-1} + T_{i,R}T_{j,R}T_{k,R}g^{-1}dg]\\ \nonumber
   &+& 5\;  d \Theta^{i}\cdot \Theta^{j}\Theta^{k}\mathrm{Tr}[(T_{i,R}T_{j,R} + T_{j,R}T_{i,R})g^{-1}T_{k,L}g] \\ \nonumber
   &-& 5\; d\Theta^{i}\cdot \Theta^{j}\Theta^{k} \mathrm{Tr}[(T_{i,L}T_{j,L} + T_{j,L}T_{i,L})g T_{k,R}g^{-1}] \\ \nonumber
   &+& 5\; \Theta^{i}\Theta^{j}\Theta^{k} \mathrm{Tr}[T_{i,L}g T_{j,R}g^{-1} T_{k,L}dg g^{-1} +T_{i,R}g^{-1} T_{j,L}g T_{k,R}g^{-1}dg]\\ \nonumber
   &+& 5\;  \Theta^{i}\Theta^{j}\Theta^{k}\Theta^{l} \mathrm{Tr}[T_{i,R}T_{j,R}T_{k,R} g^{-1}T_{l,L}g - T_{i,L}T_{j,L}T_{k,L}g T_{l,R}g^{-1}] \\
   &+& \frac{5}{2}\; \Theta^{i} \Theta^{j} \Theta^{k} \Theta^{l} \mathrm{Tr}[g T_{i,R} g^{-1} T_{j,L}g T_{k,R} g^{-1} T_{l,L}].
\end{eqnarray}
One can take the pull-back of this expression under a local section $g_{\alpha} \in \Gamma(U_{\alpha},``P")$, con $U_{\alpha} \subset {\bf D}$\footnote{Actually $``P"$ is not a $(G_L \times G_R)$-principal bundle since the action is not free. Thus we proceed formally and assume the existence of the Chern-Weil homomorphism $\mathcal{W}(\mathfrak{g}_{L} \oplus \mathfrak{g}_{R}) \to \Omega^{\ast}(``P")$ defined by mapping $\Theta^{i} T_{i,L}$ into the left connection $\omega_{L}$, $\Theta^{i} T_{i,R}$ into $\omega_{R}$ and  $u^{i} T_{i,L}$ into $F_{L}$ and $u^{i} T_{i,R}$ into $F_{R}$.}. Taking into account that $g^{\ast} \omega_{L,R} = A_{L,R}$ and $g^{\ast} \omega_{L,R} = F_{L,R}$. For a local section $g_{\alpha}$ we have $g_{\alpha}^{\ast}\omega = A_{\alpha}$, where $A_{\alpha}$ is finally the gauge potential. The gauge invariant extension for the WZW term reads
$$
\Gamma_{WZW}(g,A_{L},A_{R})= \int_{\bf D} g^{\ast} \omega^{W}_{5}
$$
$$
= \int_{\bf D} g^{\ast} \widetilde{\omega}_{5} - \sum_i \int_{\bf D} A^{i}
\wedge g^{\ast}(i_{{X_{i}}}\widetilde{\omega}_{5})
- \frac{1}{2} \sum_{i,j} \int_{\bf D}  A^{i} \wedge A^{j} \wedge  g^{\ast}
(i_{{X_{i}}} i_{{X_{j}}} \widetilde{\omega}_{5})
$$
$$
+  \frac{1}{6} \sum_{i,j,k} \int_{\bf D}  A^{i} \wedge A^{j} \wedge A^k \wedge  g^{\ast}
 (i_{{X_{i}}} i_{{X_{j}}} i_{{X_{k}}}\widetilde{\omega}_{5})
 +  \frac{1}{24} \sum_{i,j,k,l} \int_{\bf D}  A^{i} \wedge A^{j} \wedge A^k
 \wedge A^l  \wedge  g^{\ast} (i_{{X_{i}}} i_{{X_{j}}} i_{{X_{k}}}i_{{X_{l}}}\widetilde{\omega}_{5}).
$$
\begin{equation}
- {1 \over 120} \sum_{i,j,k,l,m} \int_{\bf D}  A^{i} \wedge A^{j} \wedge A^k
 \wedge A^l   \wedge A^m \wedge  g^{\ast} (i_{{X_{i}}} i_{{X_{j}}}
i_{{X_{k}}}i_{{X_{l}}} i_{{X_{m}}}\widetilde{\omega}_{5}).
\end{equation}

Making the contractions we finally found that the gauge invariant extension can be written as
\begin{eqnarray}
\label{wzw norma 5}
\nonumber
   \Gamma_{WZW}(g,A_{L},A_{R})&=& \int_{\bf D} g^{\ast}\omega_{5} + \int_{M} \bigg(-5\; \mathrm{Tr}[A_{L}\; (dg g^{-1})^{3} + A_{R}\; (g^{-1} dg)^{3} ] \\ \nonumber
   &+& 5\;  \mathrm{Tr}[(dA_{L} A_{L}+ A_{L}dA_{L}) dg g^{-1}]\\ \nonumber
   &+& 5\;  \mathrm{Tr}[(dA_{R}A_{R}+A_{R}dA_{R})g^{-1}dg]\\ \nonumber
   &+& 5\;  \mathrm{Tr}[dA_{L}dg A_{R}g^{-1}-dA_{R} dg^{-1} A_{L}g]\\ \nonumber
   &-& 5\;  \mathrm{Tr}[A_{R}g^{-1}A_{L}g (g^{-1} dg)^{2}- A_{L}g A_{R}g^{-1}(dg g^{-1})^{2}]\\ \nonumber
   &+&\frac{5}{2}\;  \mathrm{Tr}[A_{L}dgg^{-1}\wedge A_{L}dg g^{-1} -A_{R}g^{-1}dg \wedge A_{R}g^{-1}dg] \\ \nonumber
   &+& 5\;  \mathrm{Tr}[A_{L}A_{L}A_{L}dg g^{-1} + A_{R}A_{R}A_{R}g^{-1}dg]\\ \nonumber
   &+& 5\;  \mathrm{Tr}[(dA_{R}A_{R} + A_{R}dA_{R})g^{-1}A_{L}g] \\ \nonumber
   &-& 5\;  \mathrm{Tr}[(dA_{L}A_{L} + A_{L}dA_{L})g A_{R}g^{-1}] \\ \nonumber
   &+& 5\;  \mathrm{Tr}[A_{L}g A_{R}g^{-1} A_{L}dg g^{-1} +A_{R}g^{-1} A_{L}g A_{R}g^{-1}dg]\\ \nonumber
   &+& 5\;  \mathrm{Tr}[A_{R}A_{R}A_{R} g^{-1}A_{L}g - A_{L}A_{L}A_{L}g A_{R}g^{-1}] \\
   &+& \frac{5}{2}\;  \mathrm{Tr}[g A_{R} g^{-1} A_{L}g A_{R} g^{-1} A_{L}]\bigg).
\label{gieqn}
\end{eqnarray}
This result obtained through equivariant cohomology coincides with that obtained in Refs.
\cite{Witten:1983tw,Kaymakcalan:1983qq} by trial and error. More systematic constructions using geometric approaches have been proposed in Refs. \cite{Petersen:1984jy,Zumino:1983rz,Manohar:1984uq,Chou:1983qy,AlvarezGaume:1984dr,Manes:1984gk}. Thus, in general
in any dimension (including the two-dimensional case discussed in \cite{Witten:1991mm}) the equivariant
cohomology provides a tool to find in a systematic way the gauge invariant extension of the WZW term. It is remarkable that this approach using equivariant cohomology is explicitly independent on the renormalization scheme of the regularized actions, the representation of the symmetry groups and from the spacetime aspects.

Up to here we have found some nice features as: the anomaly cancelation condition and the gauge invariant extension of the WZW term can be regarded as the condition of a closed equivariant extension in the formalism of equivariant cohomology. Cartan and Weil models contain all information on these features. We will see that in addition it contains more relevant information of physical interest.

We assume that the anomalies are not canceled, i.e. the term $\mathrm{Tr}[\{T_{i,L},T_{j,L}\}T_{k,L}]-\mathrm{Tr}[\{T_{i,R},T_{j,R}\}T_{k,R}] \neq 0$, then the equivariant extension will be not closed. In this case we apply the Mathai-Quillen isomorphism to $``\widetilde{\omega}_{5}"$ and we find
\begin{equation}
\Phi_{5}^{W}= \mathrm{Tr} (g^{-1} dg)^{5} + d \Xi + 10  (\xi^{L}_{5} - \xi^{R}_{5}),
\end{equation}
where the terms $\xi_{L,R}$ have the same analytic form for the left and right parts and they have the form
\begin{equation}
\label{pre C-S}
  \xi_{5} =u^{a}u^{b} \Theta^{i}\; \mathrm{Tr}(T_{a}T_{b}T_{i}) - \frac{1}{2} \;u^{a}\Theta^{i}\Theta^{j}\Theta^{k} \;\mathrm{Tr}(T_{a}T_{i}T_{j}T_{k})+ \frac{1}{10}\; \Theta^{i}\Theta^{j}\Theta^{k}\Theta^{l}\Theta^{m} \; \mathrm{Tr}(T_{i}T_{j}T_{k}T_{l}T_{m}).
\end{equation}
After some straightforward computations we find that eq. (\ref{weil ext 5 theta}) reads
\begin{equation}
\label{weil ext nn 5}
g^{\ast}\Phi_{5}^{W}= g^{\ast}\omega_{5}^{W} + 10\;(cs^{L}_{5} - cs^{R}_{5}),
\end{equation}
where $cs_{5}^{L,R}$ are the left and right Chern-Simons five-form
\begin{equation}
cs_{5}= \mathrm{Tr}\bigg(F^{2}A- \frac{1}{2} F A^{3} + \frac{1}{10} A^{5} \bigg).
\end{equation}

It is interesting to note that the form $\xi_{5}$ defined over ${\cal W}(\mathfrak{g})\otimes\Omega^{\ast}(G)$ come from precisely the contribution of the terms  $i_{{X}}\alpha$, $i_{{X}}^{3}\beta$ y $i_{{X}}^{5} \omega_{5}$. After applying the Chern-Weil homomorphism and the pull-back under $g$ it descends to the Chern-Simons five-form $cs_{5}$ in $\Omega^{\ast}(B^{5})$. The Chern-Simons five-form $cs_5$ leads to the abelian anomaly through
the descendent Stora-Zumino procedure \cite{Zumino:1983ew} from six dimensions. The later is related to the
non-abelian anomaly of a gauge theory in four-dimensions regaining the Bardeen result
\cite{Bardeen:1969md}. The relation is performed also through the descendent procedure. The result is valid for a WZW theory defined by a Cartan $(2n+1)$-form, the generalization is immediate and direct. Of course the anomaly cancelation condition implies that $\xi_{5}$ vanishes since one can prove that the terms  $u^{a}u^{b} \Theta^{i}\; \mathrm{Tr}(T_{a}T_{b}T_{i})$, $u^{a}\Theta^{i}\Theta^{j}\Theta^{k} \;\mathrm{Tr}(T_{a}T_{i}T_{j}T_{k})$ y $\Theta^{i}\Theta^{j}\Theta^{k}\Theta^{l}\Theta^{m} \; \mathrm{Tr}(T_{i}T_{j}T_{k}T_{l}T_{m})$ separately are proportional to the anomaly cancelation condition. Consequently one has a closed equivariant extension of the WZW term and
\begin{equation}
g^{\ast}\Phi_{W}^{5}= g^{\ast}\omega_{5}^{W},
\end{equation}
with $cs^{L}_{5} - cs^{R}_{5}=0$. This condition of anomaly cancelation is not new and it can be found in the literature, see for instance, Refs. \cite{Hull:1990ms,Chou:1983qy}. However we derive it from the equivariant cohomology as a byproduct. These considerations connecting the abelian anomaly with the form $g^{\ast}\Phi_{W}^{5}$ is also valid in another spacetime dimensions. For instance in two dimensions the form of interest is the Chern-Simons 3-form emerging from $\xi^{L,R}_3$ which given by
\begin{equation}
\xi_{3}= u^{a}\Theta^{i} \mathrm{Tr}(T_{a} T_{i}) - \frac{1}{3}\; \Theta^{i}\Theta^{j}\Theta^{k} \mathrm{Tr}( T_{i}T_{j}T_{k}),
\end{equation}
which it come from the terms $i_{{X_{i}}}\lambda_{a}$ y $i_{{X_{i}}}i_{{X_{j}}}i_{{X_{k}}}\omega_{3}$ in the Weil formalism. As in the previous case if we implement the anomaly cancelation condition we get $\mathrm{Tr}(T_{a,L}T_{b,L})-\mathrm{Tr}(T_{a,R}T_{b,R})=0$ then the term $\xi_{3}^{L}-\xi_{3}^{R}$ vanishes and therefore the Chern-Simons 3-form $cs_L-cs_R$ also vanishes.

Thus it is interesting to note that starting from the Cartan 5-form and its non-closed extension in the Weil formalism it emerges in a natural way the Chern-Simons 5-form. One would ask the converse starting from the Chern-Simons 5-form and looking to recover the WZW term together with its gauge invariant extension. The answer is affirmative and one can show that the Chern-Simons term associated to a Yang-Mills theory in five dimensions plus the boundary terms codify all the relevant information of the WZW term and its gauged extension in four dimensions \cite{Hill:2006wu,Hill:2006ei}.

Finally we consider a WZW term with a left-free action in four dimensions. It defines an honest $G$-principal bundle $P$ with a natural left-free action of $G$ over $P$. Then we have a Cartan's 5-form and if we are interested in its closed equivariant extension it is given by
\begin{equation}
\label{}
{\sigma}_{5} = \omega_{5} + \sum_{a}\phi^{a} \gamma_{a} + \sum_{a,b} \phi^{a} \phi^{b} \eta_{ab},
\end{equation}
where ${\sigma}_{5}\in \Omega^{5}_{H}(G)$, $H \subset G$ is the anomaly free subgroup determined by the anomaly free condition. One can prove that such a extension of $\omega_{5}$ does exists and it is given by
\begin{eqnarray}
  \gamma_a &=& -5 {\rm Tr}[T_{a} (dg g^{-1})^{3} ], \\
  \eta_{ab} &=& 5 {\rm Tr} [ \{ T_{a},T_{b} \} dg g^{-1} ], \\
 \sum_{a,b,c} \phi^a \phi^b \phi^c i_{{X_{c}}} \eta_{ab} &=& -\sum_{a,b,c}\phi^a \phi^b \phi^c 5 {\rm Tr}[ \{ T_{a},T_{b}\} T_{c}]=0.
\end{eqnarray}
Then finally the condition $\sum_{a,b,c} \phi^a \phi^b \phi^c i_{{X_{c}}} \eta_{ab}=0$ turns out into the anomaly cancelation condition
\begin{equation}
\label{}
{\rm Tr}[\{ T_{a},T_{b}\} T_{c}]=0.
\end{equation}
This defines the anomaly free subgroup $H\subset G$, therefore we have a closed equivariant extension of $\omega_{5}$ which is precisely an element of the fifth
cohomology group of the coset space $G/H$ \cite{D'Hoker:1994ti,deAzcarraga:1998bu}.

\vskip 1truecm
%%%%%%%%%%%%%%%%%%%%%%%%%%%%%%%%%%%%%%%%%%%%%%%%%%%%%%%%%%%%%%%%%%%%%%%%%%%%%%%%%%%%%%
%%%%%%%%%%%%%%%%%%%%%%%%%%%%%%%%%%%%%%%%%%%%%%%%%%%%%%%%%%%%%%%%%%%%%%%%%%%%%%%%%%%%%%
%%%%%%%%%%%%%%%%%%%%%%%%%%%%%%%%%%%%%%%%%%%%%%%%%%%%%%%%%%%%%%%%%%%%%%%%%%%%%%%%%%%%%%
\centerline{\bf Acknoledgements}
It is a pleasure to thank E. Lupercio and B. Uribe for their advice in the subject considered in this paper. P.P. specially thanks E. Lupercio for educating him about algebraic and differential topology along the years.  The work of P.P. was supported in part by CONACyT  graduate fellowship.

%\vskip 1truecm
%%%%%%%%%%%%%%%%%%%%%%%%%%%%%%%%%%%%%%%%%%%%%%%%%%%%%%%%%%%%%%%%%%%%%%%%%%%%%%%%%%%%

\bibliography{octaviostrings}

\begin{thebibliography}{99}

\bibitem{WessZumino71}
 J. Wess, J. and B. Zumino,
  ``Consequences of Anomalous Ward Identities,''
  Phys. Lett. B {\bf 37} 95-97 (1971).

\bibitem{Witten:1983tw}
  E.~Witten,
  ``Global Aspects Of Current Algebra,''
  Nucl.\ Phys.\  B {\bf 223}, 422 (1983).
  %%CITATION = NUPHA,B223,422;%%

\bibitem{Treiman:1986ep}
  S.~B.~Treiman, E.~Witten, R.~Jackiw and B.~Zumino,
  ``Current Algebra and Anomalies,''
%\href{http://www.slac.stanford.edu/spires/find/hep/www?irn=1544780}{SPIRES entry}
{\it  Singapore, Singapore: World Scientific ( 1985) 537p}.

\bibitem{Witten:1983tx}
  E.~Witten,
  ``Current Algebra, Baryons, And Quark Confinement,''
  Nucl.\ Phys.\  B {\bf 223}, 433 (1983).
  %%CITATION = NUPHA,B223,433;%%

\bibitem{Hill:2007nz}
  C.~T.~Hill and R.~J.~Hill,
  %``Topological Physics of Little Higgs Bosons,''
  Phys.\ Rev.\  D {\bf 75}, 115009 (2007)
  [arXiv:hep-ph/0701044].
  %%CITATION = PHRVA,D75,115009;%%

\bibitem{Harvey:2007ca}
  J.~A.~Harvey, C.~T.~Hill and R.~J.~Hill,
  %``Standard Model Gauging of the Wess-Zumino-Witten Term: Anomalies, Global
  %Currents and pseudo-Chern-Simons Interactions,''
  Phys.\ Rev.\  D {\bf 77}, 085017 (2008)
  [arXiv:0712.1230 [hep-th]].
  %%CITATION = PHRVA,D77,085017;%%

\bibitem{Hill:2009wp}
  R.~J.~Hill,
  %``SU(3)/SU(2): the simplest Wess-Zumino-Witten term,''
  Phys.\ Rev.\  D {\bf 81}, 065032 (2010)
  [arXiv:0910.3680 [hep-th]].
  %%CITATION = PHRVA,D81,065032;%%

\bibitem{Witten:1991mm}
  E.~Witten,
  %``On Holomorphic Factorization Of WZW And Coset Models,''
  Commun.\ Math.\ Phys.\  {\bf 144}, 189 (1992).
  %%CITATION = CMPHA,144,189;%%

\bibitem{FigueroaO'Farrill:1994dj}
  J.~M.~Figueroa-O'Farrill and S.~Stanciu,
  %``Gauged Wess-Zumino terms and equivariant cohomology,''
  Phys.\ Lett.\  B {\bf 341}, 153 (1994)
  [arXiv:hep-th/9407196];
  ``Equivariant cohomology and gauged bosonic sigma models,''
  arXiv:hep-th/9407149.
  %%CITATION = HEP-TH/9407149;%%

%\cite{Atiyah:1984px}
\bibitem{Atiyah:1984px}
  M.~F.~Atiyah and R.~Bott,
  %``The Moment Map And Equivariant Cohomology,''
  Topology {\bf 23}, 1 (1984).
  %%CITATION = TPLGA,23,1;%%

%\cite{Mathai:1986tc}
\bibitem{Mathai:1986tc}
  V.~Mathai and D.~G.~Quillen,
  %``Superconnections, Thom classes and equivariant differential forms,''
  Topology {\bf 25}, 85 (1986).
  %%CITATION = TPLGA,25,85;%%

\bibitem{BottTuequiv}
  R. Bott and L.W. Tu,
  ``Equivariant Characteristic Classes in the Cartan Model,''
  [arXiv:math/0102001[math.DG]].

\bibitem{Wu:1993iia}
  S.~y.~Wu,
  %``Cohomological obstructions to the equivariant extension of closed invariant
  %forms,''
  J. Geom. Phys. {\bf 10}, 381-392 (1993).
  %%CITATION = PRINT-93-0083-COLUMBIA-;%%

\bibitem{Kaymakcalan:1983qq}
  O.~Kaymakcalan, S.~Rajeev and J.~Schechter,
  %``Nonabelian Anomaly And Vector Meson Decays,''
  Phys.\ Rev.\  D {\bf 30}, 594 (1984).
  %%CITATION = PHRVA,D30,594;%%

%\cite{Petersen:1984jy}
\bibitem{Petersen:1984jy}
  J.~L.~Petersen,
  %``Nonabelian Chiral Anomalies And Wess-Zumino Effective Actions,''
  Acta Phys.\ Polon.\  B {\bf 16}, 271 (1985).
  %%CITATION = APPOA,B16,271;%%

%\cite{Zumino:1983rz}
\bibitem{Zumino:1983rz}
  B.~Zumino, Y.~S.~Wu and A.~Zee,
  %``Chiral Anomalies, Higher Dimensions, And Differential Geometry,''
  Nucl.\ Phys.\  B {\bf 239}, 477 (1984).
  %%CITATION = NUPHA,B239,477;%%

%\cite{Manohar:1984uq}
\bibitem{Manohar:1984uq}
  A.~Manohar and G.~W.~Moore,
  %``Anomalous Inequivalence Of Phenomenological Theories,''
  Nucl.\ Phys.\  B {\bf 243}, 55 (1984).
  %%CITATION = NUPHA,B243,55;%%

%\cite{Chou:1983qy}
\bibitem{Chou:1983qy}
  K.~c.~Chou, H.~y.~Guo, K.~Wu and X.~c.~Song,
  %``On The Gauge Invariance And Anomaly Free Condition Of Wess-Zumino-Witten
  %Effective Action,''
  Phys.\ Lett.\  B {\bf 134}, 67 (1984).
  %%CITATION = PHLTA,B134,67;%%

%\cite{AlvarezGaume:1984dr}
\bibitem{AlvarezGaume:1984dr}
  L.~Alvarez-Gaume and P.~H.~Ginsparg,
  %``The Structure Of Gauge And Gravitational Anomalies,''
  Annals Phys.\  {\bf 161}, 423 (1985)
  [Erratum-ibid.\  {\bf 171}, 233 (1986)].
  %%CITATION = APNYA,161,423;%%

\bibitem{Manes:1984gk}
  J.~L.~Manes,
  %``Differential Geometric Construction Of The Gauged Wess-Zumino Action,''
  Nucl.\ Phys.\  B {\bf 250}, 369 (1985).
  %%CITATION = NUPHA,B250,369;%%

\bibitem{Zumino:1983ew}
  B.~Zumino,
  ``Chiral Anomalies And Differential Geometry'', B.S. De Witt and R. Stora, eds. Les Hauches, Session XL, 1983, {\it Relativity, groups and topology II}, Elsevier Science Publishers B.V., 1984.

\bibitem{Bardeen:1969md}
  W.~A.~Bardeen,
  %``Anomalous Ward identities in spinor field theories,''
  Phys.\ Rev.\  {\bf 184}, 1848 (1969).
  %%CITATION = PHRVA,184,1848;%%

%\cite{Hull:1990ms}
\bibitem{Hull:1990ms}
  C.~M.~Hull and B.~J.~Spence,
  %``The Geometry of the gauged sigma model with Wess-Zumino term,''
  Nucl.\ Phys.\  B {\bf 353}, 379 (1991).
  %%CITATION = NUPHA,B353,379;%%

%\cite{Hill:2006wu}
\bibitem{Hill:2006wu}
  C.~T.~Hill,
  %``Exact equivalence of the D = 4 gauged Wess-Zumino-Witten term and the D  =
  %5 Yang-Mills Chern-Simons term,''
  Phys.\ Rev.\  D {\bf 73}, 126009 (2006)
  [arXiv:hep-th/0603060].
  %%CITATION = PHRVA,D73,126009;%%

%\cite{Hill:2006ei}
\bibitem{Hill:2006ei}
  C.~T.~Hill,
  %``Anomalies, Chern-Simons terms and chiral delocalization in extra
  %dimensions,''
  Phys.\ Rev.\  D {\bf 73}, 085001 (2006)
  [arXiv:hep-th/0601154].
  %%CITATION = PHRVA,D73,085001;%%

%\cite{D'Hoker:1994ti}
\bibitem{D'Hoker:1994ti}
  E.~D'Hoker and S.~Weinberg,
  %``General effective actions,''
  Phys.\ Rev.\  D {\bf 50}, 6050 (1994)
  [arXiv:hep-ph/9409402].
  %%CITATION = PHRVA,D50,6050;%%

%\cite{deAzcarraga:1998bu}
\bibitem{deAzcarraga:1998bu}
  J.~A.~de Azcarraga and J.~C.~Perez Bueno,
  %``On the general structure of gauged Wess-Zumino-Witten terms,''
  Nucl.\ Phys.\  B {\bf 534}, 653 (1998)
  [arXiv:hep-th/9802192].
  %%CITATION = NUPHA,B534,653;%%



\end{thebibliography}
\addcontentsline{toc}{section}{Bibliography}
\bibliographystyle{TitleAndArxiv}

%\end{document}%
%---------------------------------------------------------------------------------------
%                       References
%----------------------------------------------------------------------------------------

\end{document}